\documentclass[12pt]{iopart}

\pdfoutput=1
\usepackage{amssymb}
\usepackage{iopams}
\usepackage{setstack}
\usepackage{amsthm}
\usepackage{graphicx}
\usepackage[all]{xy}
\usepackage{color}
\usepackage{subfig}



\newcommand{\blue}[1]{\textcolor{blue}{#1}}

\newcommand{\beq}{\begin{equation}}
\newcommand{\eeq}{\end{equation}}
\newcommand{\beqa}{\begin{eqnarray}}
\newcommand{\eeqa}{\end{eqnarray}}

\def \P {{\mathcal P}}

\newcommand\ket[1]{\ensuremath{
    \vert{#1}\mkern1.2mu\rangle}}
\newcommand\bra[1]{\ensuremath{
    \langle{#1}\mkern1.2mu\vert}}

\theoremstyle{definition}

\begin{document}
\title{Ultracold Chemistry and its Reaction Kinetics}

\author{Florian Richter}
\address{Institute for Theoretical Physics, Leibniz Universit\"at Hannover,
Appelstra\ss e 2, 30167 Hannover, Germany}
\author{Daniel Becker}
\address{Departement of Physics, Universit\"at Basel,
Klingelbergstrasse 82, CH-4056 Basel, Switzerland}
\author{C\'edric B\'eny}
\address{Institute for Theoretical Physics, Leibniz Universit\"at Hannover,
Appelstra\ss e 2, 30167 Hannover, Germany}
\author{Torben A. Schulze}
\address{Institute for Quantum Optics, Leibniz Universit\"at Hannover,
Welfengarten 1, 30167 Hannover, Germany}
\author{Silke Ospelkaus}
\address{Institute for Quantum Optics, Leibniz Universit\"at Hannover,
Welfengarten 1, 30167 Hannover, Germany}
\author{Tobias J. Osborne}
\address{Institute for Theoretical Physics, Leibniz Universit\"at Hannover,
Appelstra\ss e 2, 30167 Hannover, Germany}

\date{\today}
\begin{abstract}
We study the reaction kinetics of chemical processes occurring in the ultracold regime and systematically investigate their dynamics. Quantum entanglement is found to play a key role in driving an ultracold reaction towards a dynamical equilibrium. In case of multiple concurrent reactions Hamiltonian chaos dominates the phase space dynamics in the mean field approximation.
\end{abstract}


\maketitle
\clearpage


\section{Introduction}
A chemical reaction normally occurs at a few hundred kelvin between reagents involving large numbers of particles ($\sim 10^{23}$). This is because reactions are usually activated by \emph{thermal fluctuations} which are only significant for large concentrations of particles with high momenta. Therefore, the possibility of a chemical reaction taking place in the dilute and ultracold ($T<1\, \mu $K) regime is somewhat counterintuitive.
However, the formation of ultracold molecules from atoms in a Bose-Einstein condensate (BEC) was observed almost 15 years ago \cite{wynar2000molecules}.
This interaction of atoms and molecules close to the absolute zero of temperature has been referred to as \emph{ultracold chemistry} \cite{hutson2010ultracold}.


A variety of experimental techniques such as the coupling of atoms and molecules via magnetic Feshbach resonance \cite{donley2002atom,regal2003creation} have been successfully employed to achieve chemical bonding in an ultracold environment. For example, ultracold Potassium-Rubidium molecules \cite{ospelkaus2010quantum}  have been investigated to analyse the quantum mechanical effects of particle statistics on molecular reactivity. The now ready experimental accessibility of chemical processes in the dilute ultracold regime strongly motivates us to develop a general physical understanding of their \emph{reaction kinetics}.

A quantized description is required in order to study the dynamics of ultracold reactions, in order to fully account for the effects of quantum fluctuations and entanglement. Here we should replace the classical notion of a temperature-dependent reaction with a coherent reversible Hamiltonian evolution. The first phenomenological steps toward such a description were taken in \cite{heinzen2000superchemistry}, where a mean field ansatz was exploited to describe the coherent formation of diatomic molecules in a BEC. Since then a variety of extensions to this model, mainly focussed on adding quantum corrections to the original mean field ansatz, have been studied \cite{goral2001multimode,Santos2005classical}. Concomitantly, theoretical explanations, via two-body scattering processes, for the ultracold chemical reaction rates observed in recent experiments have been proposed \cite{carr2009cold}. In the case of fermionic particles with several internal states the observed reaction rates may be understood in terms of a simple quantum threshold model \cite{quemener2010strong}. When there are different particle types, e.g.\ bosons and fermions, more sophisticated modelling in terms of multichannel quantum defect theory \cite{idziaszek2010universal} is required. So far, a general and systematic investigation of the dynamics of ultracold reactions, analogous to the study of reaction kinetics for classical thermal reactions, has not yet been undertaken.
Such an approach seems to be indispensable if one wants to study the role of quantum coherence and the production of quantum entanglement in ultracold chemical systems.


In this paper we propose a general scheme to study the kinetics of ultracold chemical reactions. Investigating the predicted dynamics, we find that entanglement replaces the role of thermal fluctuations in activating ultracold chemical reactions. This allows us to draw a parallel between ultracold chemical reactions and quenched dynamics. Thanks to the generality of our formulation it is possible to consider complex chemical reactions involving many reagents. We exemplify our approach by studying an experimentally accessible example of an ultracold reaction exhibiting rich quantum phenomena leading to a dynamical relaxation to local equilibrium.

\section{Classical Reaction Kintetics}
The subject of reaction kinetics in classical chemistry is concerned with the temporal dynamics of a chemical reaction and its reaction rate. One usually studies a general reaction
\begin{equation}
\sum_{i} \mu_i A_{i} \overset{k_{ab}}{\underset{k_{ba}}{\rightleftarrows}} \sum_{j} \nu_j B_{j},
\label{eq:reaction_1}
\end{equation}
where the numbers $\mu_i$ and $\nu_j$ are referred to as \emph{stoichiometric coefficients} and $k_{ab}$ and $k_{ba}$ are called \emph{reaction constants}. If the reaction is elementary, i.e.\ there are no catalytic, intermediate, or concurrent reactions, the rate equation for the concentration of species $A_i$, denoted as $[A_i]$, can be inferred straightforwardly \cite{connors1990chemical}:
\begin{equation}
\frac{1}{\mu_i}\frac{d[A_i]}{dt} =-k_{ab}\prod_j [A_j]^{\mu_j}+k_{ba}\prod_k [B_k]^{\nu_k}.
\label{eq:reaction_2}
\end{equation}
In the terminology of irreversible reactions, i.e.\ when either $k_{ab}$ or $k_{ba}$ equals zero, the number $\max\{M,N\}$, where $M:=\sum_j \mu_j$ and  $N:=\sum_k \nu_k$, is called the \emph{order} of the reaction. Systems of parallel reactions are treated in an analogous manner. An important feature of classical reaction kinetics is that the temperature dependence of the reaction constants is described by Arrhenius' law:
\begin{equation}
k(T)\propto\exp\left(\frac{-E_A}{\kappa T}\right),
\end{equation}
where $E_A$ represents the \emph{activation energy} of the corresponding reaction. Microscopically this means the statistical fluctuations drive the reaction.
This model of classical chemical reaction kinetics has been validated by its long term success in describing a wide range of astonishing phenomena from the oscillating Belousov-Zhabotinskii reaction \cite{belousov1959periodic,zhabotinsky1964periodical} to deterministic chaos \cite{epstein1996nonlinear}. However, classical reaction kinetics is still an active field, e.g., the question of the existence of an equilibrium was only recently answered in \cite{baez2012course}.

\section{Proposed Framework}
The description of the dynamics of chemical reactions at ultracold temperatures falls into the framework of quantum mechanics as the basic assumption of Arrhenuis' Law, namely a disordered movement of many high-momentum particles, is no longer satisfied in the low temperature regime. Hence we must replace Arrhenius' law with a new concept. Here we propose that \emph{quantum entanglement} provides the driving source of fluctuations in the ultracold regime. This is because when the system is in a non-equilibrium pure state quantum correlations are generically induced by unitary dynamics \footnote{Unless the dynamics is trivial, i.e., noninteracting, or the initial state is an equilibrium pure state.}. An analogous situation occurs in the study of \emph{quench dynamics} for quantum systems \cite{cramer2008exact,cramer2010quantum}, i.e.\ large many-body systems where the interaction undergoes a sudden change at some fixed time. Based on this analogy, we expect the dynamics of ultracold reactions to be dominated by entanglement-induced phenomena, such as the relaxation of subsystems to a local equilibrium.

A chemical reaction is inherently a many particle problem so that second quantization is the natural framework. Hence we describe every reacting species $A_i$ with a corresponding quantum field $\hat{\psi}_{A_i}$. The underlying Hilbert space is a tensor product of the single-species Fock spaces. The role of the classical particle concentration $[A]$ is replaced by the particle density operator $\hat{n}_A=\hat{\psi}_A^\dag\hat{\psi}_A$. The dynamics of the system is induced by its Hamiltonian $\hat{H}=\hat{H}_0+\hat{H}_{\text{int}}$, where $\hat{H}_0=\hat{H}_{\text{kin}}+\hat{H}_{\text{pot}}$ represents the standard second-quantised kinetic and potential terms, respectively. The crucial part of our proposal is now the choice of a proper interaction term $H_{\text{int}}$ to describe particle conversion, such as molecule formation. Taking guidance from the classical setting we propose the following $\hat{H}_{\text{int}}$ for the elementary reaction Eq.~\Eref{eq:reaction_1}:
\begin{equation}
\hat{H}_{\text{int}}=k\int{dx}\prod_{i,j}\left(\hat{\psi}_{A_i}^\dag(x)\right)^{\mu_i}\left(\hat{\psi}_{B_j}(x)\right)^{\nu_i} + \text{h.c.},
\end{equation}
where $\hat{\psi}_{A_i}$ and $\hat{\psi}_{B_j}$ obey the canonical commutation relations (CCR) or canonical anti-commutation relations (CAR) according to particle type of the species. If there are multiple concurrent reactions one should take the sum of each of the interaction Hamiltonians $\hat{H}_{\text{int}}=\sum_i\hat{H}_{\text{int},i}$. Notice that within this framework we are restricted to the  description of reversible reactions with a single reaction rate $k\equiv k_{ab}=k_{ba}$. However, with a little work, the interaction Hamiltonian can be naturally extended via the inclusion of ancillary baths to incorporate more general effects such as a self-interaction of particles, particle loss, and dissipation. For the rest of the paper, we neglect those effects. We make an additional crucial assumption, namely that the positional degrees of freedom are not excited during the reaction. Therefore, we perform a cutoff after the first term in the operator mode expansion, i.e.\ we replace field operators with annihilation operators according to $\hat{\psi}_{A_i}\rightarrow \phi^{A_i}_{0}(x) \hat{a}_{A_i}$, where $\phi^{A_i}_0(x)$ denotes the single-particle ground state wave-function of the respective species. Consequently, the density observable is replaced by the number operator $\hat{n}_{A_i}=\hat{a}^\dag_{A_i}\hat{a}_{A_i}$ and $\hat{H}_0$ simplifies to a harmonic oscillator: $\hat{H}_0=E_{A_i}\hat{n}_{A_i}$. For simplicity we also restrict ourselves to the bosonic systems although the addition of fermionic particles is straightforward. Keeping these approximations in mind we summarise in Table \ref{tab:interaction} the proposed interaction Hamiltonians for the first few elementary low-order ultracold chemical reactions.

\begin{table}[h!]
	\centering
		\begin{tabular}{c|c|c}
		 Order&Reaction & Proposed $\hat{H}_{\text{int}}$ \\ \hline
			0.&$\text{bath} \overset{k}{\rightleftarrows} A$ & $k(\hat{a}_A^\dag+ \text{h.c.})$  \\
			1.&$A \overset{k}{\rightleftarrows} B$ & $k(\hat{a}^\dag_{A} \hat{a}_{B} + \text{h.c.})$  \\
			2.&$A + B \overset{k}{\rightleftarrows} C$ & $k(\hat{a}_{A}^\dag\hat{a}_{B}^\dag\hat{a}_{C} + \text{h.c.})$  \\
			3.&$A + B \overset{k}{\rightleftarrows} C + D$ &  $k(\hat{a}_{A}^\dag\hat{a}_{B}^\dag\hat{a}_{C} \hat{a}_{D} + \text{h.c.})$
		\end{tabular}
		\vspace{3mm}
		\caption{The proposed interaction Hamiltonians for low-order bosonic reactions.}
			\label{tab:interaction}
\end{table}

To complete our proposal, we need to consider suitable initial conditions for our dynamical system. It does not make sense to initialize the system in an equilibrium state, i.e.\ an eigenstate or thermal state of $\hat{H}_0+\hat{H}_{\text{int}}$, if one is interested in dynamical effects. Rather, we assume that the system is in an eigenstate of the non-interacting part of the Hamiltionian, which corresponds to the dynamical isolation of the reacting species before the chemical reaction starts. In order to make the comparison to classical reaction kinetics as meaningful as possible, we consider the system to be initialized in a product of coherent states~\footnote{To avoid experimentally unrealistic situtations, we also restrict our considerations to states with finite particle number.}, although the generalization to different initial conditions is obvious.

Considering the scheme outlined in table \ref{tab:interaction}, we see that a zeroth-order reaction is not a chemical reaction per se, due to the absence of any interaction of different species. It can be rather understood as a connection of the considered species to some reservoir. Neglecting $H_0$, this coupling $\hat{H}_{\text{int}}=k(\hat{a}_A^\dag+\hat{a}_A)$ results in a quadratic scaling of the average particle number of the coupled species $\left<\hat{n}_A(t)\right>\propto t^2$. Notice, by way of contrast, that in the classical case we obtain a linear growth or decay of the particle concentration $\left[A\right]\propto\pm t$ according to the choice of the sign of the coupling constant.

There are two types of elementary first-order reaction. The first describes the simplest of all ultracold chemical reactions, namely a simple particle conversion between two species. The Hamiltonian $H_{\text{int}}=k(\hat{a}^\dag_{A} \hat{a}_{B} + \text{h.c.})$ models a linear interaction between two quantum fields (e.g.\ a beam-splitter in quantum optics \cite{Mar-Sarao_08}). It is exactly solvable in the sense that it can be linearly transformed into decoupled harmonic oscillators. The average particle number of each species therefore periodically oscillates, in contrast to the classical first-order reaction, which relaxes to a stationary state. The other type of first-order reaction is modelled by $H_{\text{int}}=k(\hat{a}^\dag_{A} \hat{a}_{B}^\dag + \text{h.c.})$ and describes the production of ``pairs'' from a bath. This interaction is familiar in quantum optics where it models two-mode squeezing.

In order to describe more complex chemical reactions, e.g.\ the formation of a molecule from two particles, we have to go beyond first-order reactions. The corresponding interaction Hamiltonians describe non-quadratically interacting field theories. These models are not generally exactly solvable and we must employ approximations, heuristics, and numerical methods to study their dynamics.

\section{Diatomic Molecule Formation}
Here we study to the most elementary second-order reaction, \emph{diatomic molecule formation} \cite{yurovsky2006formation}:
\begin{equation}
A + A \overset{k}{\rightleftarrows} A_2
	\label{eq:molecule}
\end{equation}
The stoichiometric coefficient for the atomic species $\mu_A$ equals two and the molecular coefficient $\nu_{A_2}$ is one. According to our proposed framework this reaction is modelled by the Hamiltonian:
\begin{equation}
\hat{H} =E_{A}\hat{n}_A +E_{A_2}\hat{n}_{A_2} + k(\hat{a}_A^\dag\hat{a}_A^\dag\hat{a}_{A_2} + h.c.),
		\label{eq:quantum_hamiltonian}
\end{equation}
where $E_A$ and $E_{A_2}$ label the corresponding ground-state energies. This Hamiltonian has been the subject of much recent research as it also models second harmonic generation \cite{walls1972nonlinear} or two-photon down conversion \cite{hillery1990photon}, as well as molecule formation via coherent photoassociation in an atomic BEC \cite{javanainen1999coherent,javainen2,santos_links}. Nonetheless, a comprehensive investigation of its dynamical regimes relevant for the reaction kinetics of ultracold chemistry is missing.

We are interested in the kinetics of the reaction \eref{eq:molecule}, and in particular the role of entanglement, so we restrict our analysis to the interacting part of the Hamiltonian and assume the reaction rate to be much larger than the ground state energies, i.e.\ $k \gg |E_A| + |E_{A_2}|$. 
Additionally, the overall particle number operator $\hat{N}_{tot}=\hat{N}_A+2\hat{N}_{A_2}$ is a conserved quantity for the dynamics, hence we can restrict our study to the reduced dynamics of the atomic mode, writing $N \equiv N_A$. The behaviour of the molecular mode can be deduced straightforwardly.

\begin{figure}[htbp]
	\centering
		\includegraphics[width=0.8\textwidth]{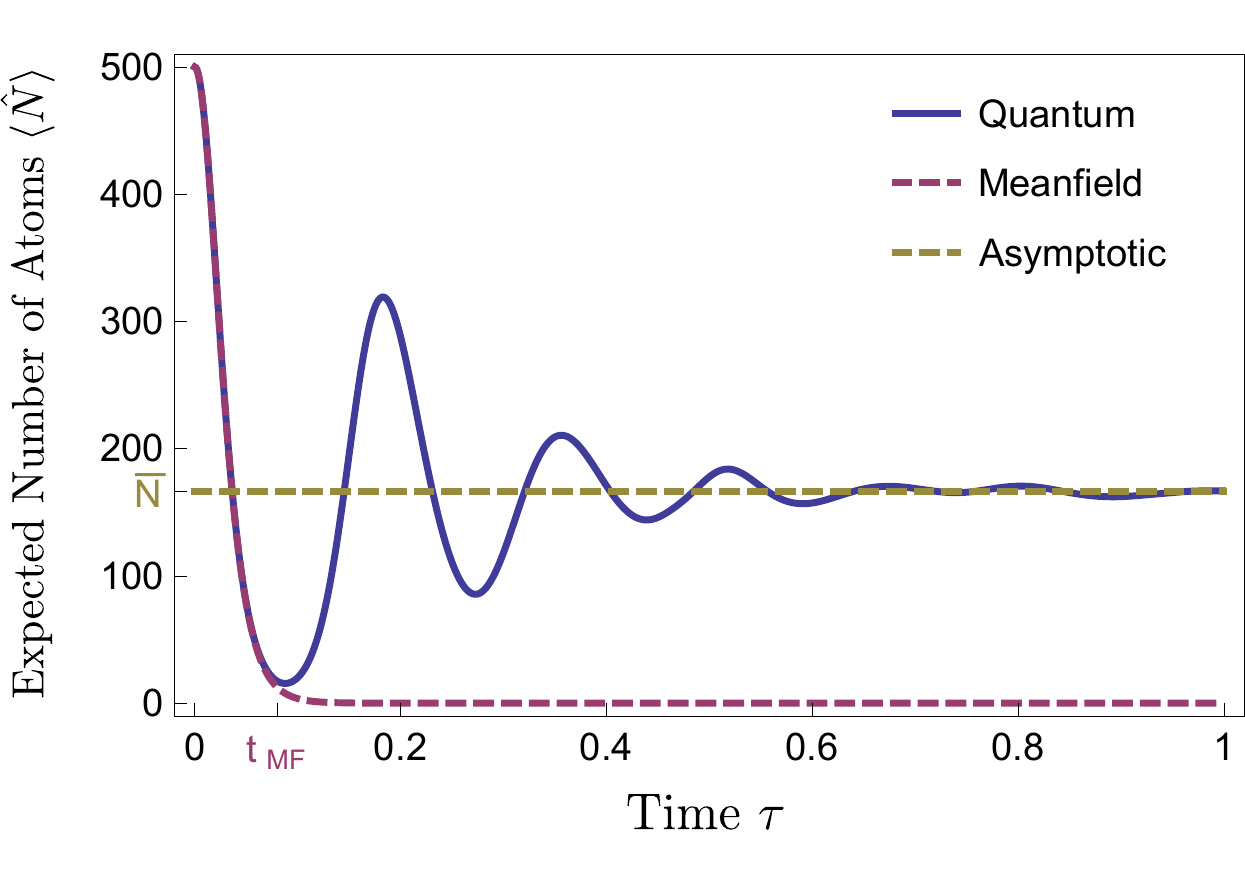}
	\caption{The dynamics of the atomic occupation number expectation value $\langle \hat{N} \rangle$ with respect to the 
	rescaled time $\tau=tk/\hbar$ for the diatomic molecular reaction $A + A \rightleftarrows A_2$, with $N=500$ particles. The quantum and mean field trajectories coincide until they separate at $\tau_{\rm MF}$. From there on the quantum trajectory approaches a stationary value $\overline{N}$ via entanglement-induced damped oscillations.}
	\label{fig:1}
\end{figure}
In Fig~\ref{fig:1} we have plotted the full quantum time evolution of the expectation value of $\hat{N}$ (obtained via exact diagonalisation) together with the mean field prediction for the dynamics of the reaction. The system is initially in a product state of a coherent atomic state and a completely depleted molecular state. Mean field theory predicts a complete inversion of the population, where the system is driven to an unstable fixed point \cite{vardi2001quantum}. However, considering the full quantum solution, we identify three different dynamical regimes: First, a \textit{semi-classical} regime, where the quantum and mean field dynamics coincide. At the \emph{breakdown time} $\tau_{\rm MF}$ the full solution drifts away from the mean field approximation \cite{vardi2001bose} and the semi-classical regime transitions to an intermediate \textit{evanescent regime}, where the quantum trajectory oscillates with an increasingly damped amplitude. Eventually, the system reaches the \textit{asymptotic} regime, where the expectation value of the population imbalance relaxes to a stationary value $\overline{N}$.

\begin{figure}[htbp]
	\centering
		\includegraphics[width=0.8\textwidth]{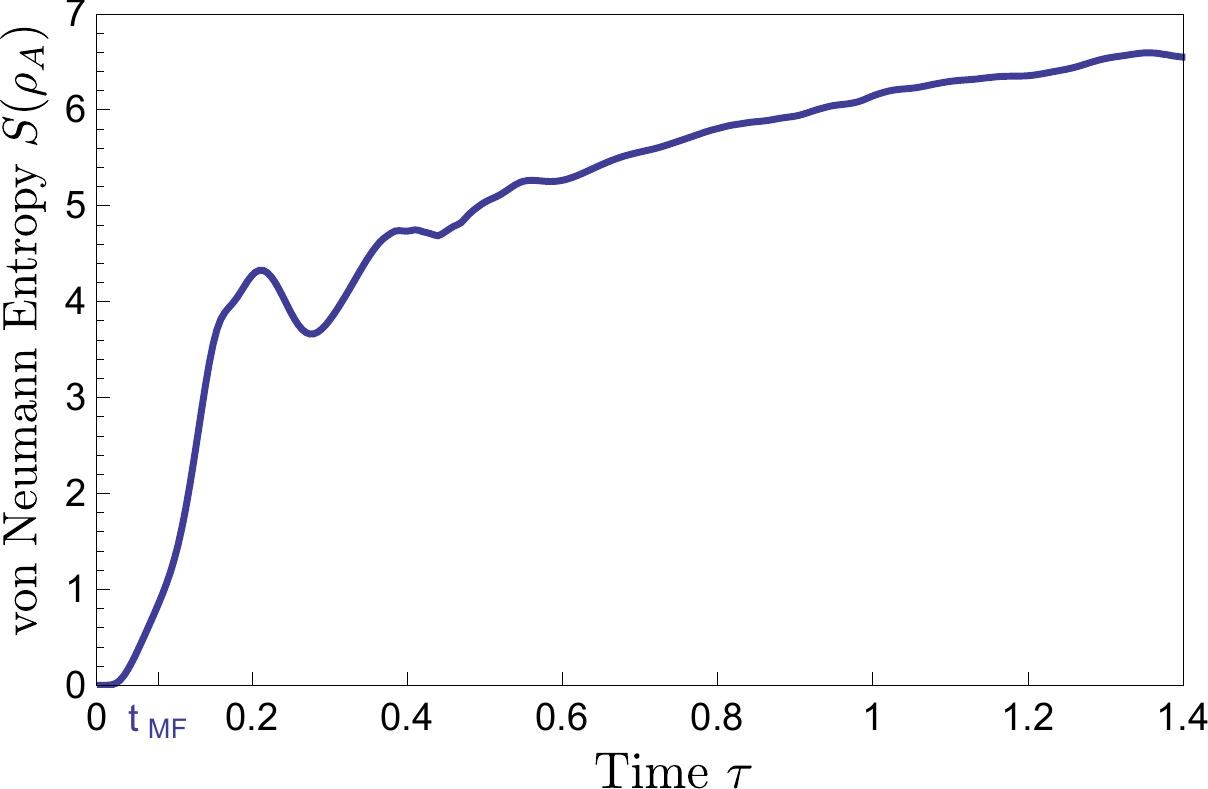}
	\caption{Evolution of the quantum entanglement (in terms of the von Neumann entropy of the atomic reduced density operator $\rho_A$) between the atomic and molecular modes for the reaction Eq.~(\ref{eq:molecule}) with respect to rescaled time $\tau=tk/\hbar$. For an initially coherent atomic state the amount of entanglement is close to zero for short times. In the vicinity of the breakdown time $t_{MF}$ the entanglement rapidly increases and stays roughly at a constant level for later times.}
	\label{fig:2}
\end{figure}
We can understand the three different dynamical regimes by studying the time evolution of the quantum entanglement between the atomic and molecular modes. These results are shown in Fig.~\ref{fig:2}. In the semi-classical regime we see a rapid increase of the entanglement at the beginning of the reaction, which is necessary for the formation of molecules. It is initially rather surprising that the mean field approximation works as well as it does in the semi-classical regime given that the state rapidly becomes entangled and is not well-modelled via a product ansatz. A possible explanation is that the entanglement evolution in the semi-classical regime is typical of that produced by \emph{integrable interactions} \cite{hines2003entanglement}, at least until the breakdown time $\tau_{\rm MF}$. After the breakdown time, in the evanescent regime, the system rapidly reaches the maximum available entanglement and begins to explore the full  Hilbert space. Soon after, it enters the asymptotic regime where it ergodically evolves through highly entangled states. It remains in the asymptotic regime until it experiences a quantum revival.

\begin{figure}[htbp]
	\centering
		\includegraphics[width=0.8\textwidth]{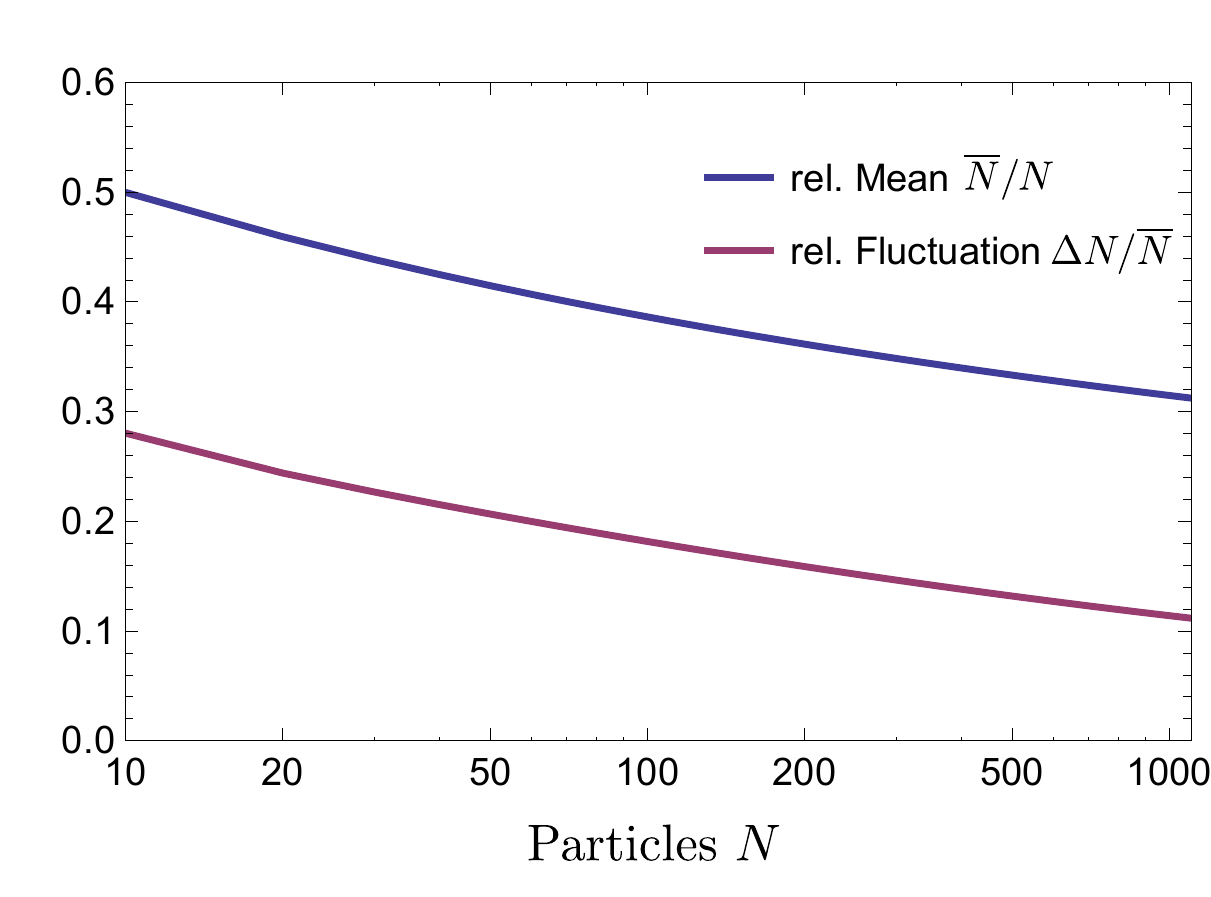}
	\caption{The time-averaged relative mean value of atoms and the fluctuations around it against the overall particle number for the reaction Eq.~(\ref{eq:molecule}). As the particle number is increased the fraction of atoms in the asymptotic regime and the relative temporal fluctuations decrease.}
		\label{fig:3}
\end{figure}

The dynamical behavior exhibited by the reaction \Eref{eq:molecule} is reminiscent of the local relaxation observed in quenched many particle quantum systems \cite{cramer2008exploring}. This hypothesis is supported by studying the the time-averaged fluctuations $\Delta N^2\!=\!\lim_{\tau \rightarrow \infty} \frac{1}{\tau}\int_0^\tau dt\left(\langle{\hat{N}(t)}\rangle-\overline{N}\right)^2$ relative to the mean value $\overline{N}\!=\!\lim_{\tau \rightarrow \infty} \frac{1}{\tau}\int_0^\tau dt\langle\hat{N}(t)\rangle$, plotted in Fig.~\ref{fig:3}: we find that the fluctuations decrease as the particle number $N$ is increased. However, the mechanism leading to the local relaxation observed in quenched dynamics is slightly different to that found here. In quenched many particle systems the incoherent interference of localised excitations travelling at different velocities leads to a cumulative effect of relaxation. However, in our case, we have an interaction between just two modes and the relaxation we observe here is directly related to the growth of entanglement between them, namely the loss of coherence, or purity, of the reduced density operators. 
Finally, we point out that the relaxation behavior in the asymptotic regime is remarkably similar to the classical high temperature kinetics of \Eref{eq:molecule}, which relaxes to the fixed point $\left[A \right]^2=\left[A_2 \right]$, even though our system is always in a global pure state. 

We obtained $\overline N$ and $\Delta N^2$ via exact diagonalisation. Although the full Hamiltonian has degenerate eigenvalues, the dynamical problem, due to the conservation of $\hat{N}_{\text{tot}}$, can be separated into finite-dimensional problems with no degeneracy. Exploiting this we find that the time-averaged expectation value of the atoms and the molecules coincides with the predictions given by the diagonal ensemble $\overline{N}_{\text{ens}}\!:=\!\sum_\alpha|c_{\alpha}|^2 N_{\alpha,\alpha}$, where $c_\alpha=\bra{\psi_{\text{in}}}\alpha\rangle$ and $N_{\alpha,\beta}=\bra{\alpha}\hat{N}\ket{\beta}$ \cite{neumann1,rigol2008thermalization} are the coefficients in the complete energy eigenbasis. Moreover, the time-averaged fluctuations around this mean value can be obtained via  $\Delta N_{t}^2\!=\!\sum_{\alpha\neq\beta}|c_{\alpha}|^2|c_{\beta}|^2 |N_{\alpha,\beta}|^2$ \cite{Srednicki}. However, the system \emph{does not thermalize} as the predicted expectation values do not coincide with those of the microcanonical ensemble.


For an experimental implementation of \Eref{eq:molecule} we consider ultracold Cesium atoms:
\begin{equation}
\mathrm{Cs} + \mathrm{Cs} \overset{k}{\rightleftarrows} \mathrm{Cs}_{2}
\end{equation}
with $\mathrm{Cs_{2}} = |^{3}\Sigma^{+}_{u}, v = 0, J = 0 \rangle$ and $\mathrm{Cs} = |F = 4, m_{F} = 4 \rangle$. The chosen atomic state has pure triplet spin character, and we neglect spin mixing effects due to Spin-Orbit coupling, therefore excluding any vibrational relaxation to the singlet ground state. Potential energy curves of the states under consideration are available through \cite{xie2009experimental,krauss1990effective,allouche2012transition}. $E_{A}$ and $E_{A_2}$ are given by an optical trap confinement and will be different and  tunable due to the different dynamic polarizabilities of the molecular and atomic state \cite{vexiau_optimal_2011}. For our calculation, $E_{M} = 1.05, E_{A} = 1.05 (\frac{1}{2}m \omega_{A}^{2})$ with $ \omega_{A} = 50\mathrm{Hz}.$ The reaction is realized through a 2-color photoassociation in continuous Raman configuration. For the molecule conversion, one has $k = \frac{k_{1} k_{2}}{2\Delta}$, where $k_{i}$ (i = 1,2) are the individual Rabi frequencies and $\Delta$ the one-photon detuning. We evaluate the corresponding wavefunctions and transition dipole matrix elements to and from the intermediate states $|b^{3}\Pi_{u}, v', J'=0 \rangle$ and $|c^{3}\Sigma_{g}^{+}, v', J'=0 \rangle $. As an example, for a transition to the $|b^{3}\Pi_{u}, v', J'=0 \rangle$ manifold via lasers operating around 1275nm and 1232nm respectively, our calculation yields Rabi frequencies $2\pi \times 1.17\mathrm{MHz} [0.4 \mathrm{kHz}] \sqrt{I/\mathrm{mW}/\mathrm{cm}^{2}}$ for the molecular [atomic] coupling respectively. Therefore for a typical Raman setup ($\Delta=750 \mathrm{MHz}, I_{1} = I_{2} = (10\mathrm{mW}/0.25 \mathrm{mm}^{2})$, hence $k = 1.25 \mathrm{kHz}$), $k > |E_A| + |E_{A_2}|$ is safely fulfilled.

\section{Concurrent Reaction}
In classical high-temperature kinetics we need to consider complex reactions involving numerous reactants in order to obtain oscillating or irregular dynamics. However, we will see that when we add to the ultracold diatomic molecule formation a simple zeroth order reaction\blue{,} we already encounter Hamiltonian chaos in the mean field regime. For this purpose consider the chemical reaction
\begin{eqnarray}
A+A \overset{k_1}{\rightleftarrows} A_2 \\ 
\text{bath} \overset{k_2}{\rightleftarrows} A
\label{eq:concurrent}
\end{eqnarray}
Applying the proposed rules and assuming that our particles are trapped in the ground state of some harmonic potential, we obtain the following Hamiltonian:
\begin{equation}
\hat{H}=E_A\hat{n}_A +E_{A_2}\hat{n}_{A_2} + k_1(\hat{a}_A^\dag\hat{a}_A^\dag\hat{a}_{A_2} +\hat{a}_A\hat{a}_A\hat{a}_{A_2}^\dag)+k_2(\hat{a}^\dag_A+\hat{a}_A),
\label{eq:hamilton_expanded}
\end{equation}
where $E_A$ and $E_{A_2}$ denote the ground state energy of the respective molecular or atomic species. The concurrent zeroth order reaction \eref{eq:concurrent} breaks the conservation of the overall particle number $\hat{N}_{tot}$ and therefore impedes a full quantum mechanical treatment. Consequently, we investigate the dynamics of the system in the mean field approximation by replacing the creation and annihilation operators in \eref{eq:hamilton_expanded} with complex numbers $(\alpha_A,\alpha_{A_2})$ labeling coherent states. Note that within this approximation we obtain the average particle number of a certain species $\langle \hat{N_i}\rangle$ by considering the square of the modulus of the respective complex number $|\alpha_i|^2$.

In the previous section we have seen that the diatomic molecule formation amounts to deviations from mean field dynamics because of the occurrence of quantum effects. Therefore, we need to keep the coupling parameter $k_1$ as small as possible compared to some relevant energies to consider the mean field limit as an appropriate description of the actual dynamics. Keeping this in mind, we obtain the equations of motion from the variational principle: 
\begin{eqnarray}\eqalign{
i\dot{\alpha}_A&=E_A\alpha_A + 2k_1\overline{\alpha}_A\alpha_{A_2}+k_2\\
i\dot{\alpha}_{A_2}&=E_{A_2}\alpha_{A_2} + k_1\alpha_A^2}.
\end{eqnarray}
To reduce the number of parameters in the system, we remove unnecessary degrees of freedom by replacing the dynamical variables with nondimensionalized quantities
\begin{eqnarray}  
\tilde{\alpha}_A&=\frac{\alpha_A}{\alpha_{0}} , \qquad \tilde{\alpha}_ {A_2}=\frac{\alpha_{A_2}}{\alpha_{0}},\qquad \tau=\frac{t}{t_0}.
\end{eqnarray}
This amounts to the coupled equations
\begin{eqnarray}\eqalign{
i\frac{d\tilde{\alpha}_A}{d\tau} - \tilde{\alpha}_A = 1 + c_12\overline{\tilde{\alpha}}_A\tilde{\alpha}_{A_2}\\
i\frac{d\tilde{\alpha}_{A_2}}{d\tau} - c_2\tilde{\alpha}_{A_2} =c_1\tilde{\alpha}_A^2,}
\label{eq:coupled}
\end{eqnarray}
and energy
\begin{equation}
\tilde{H}=|\tilde{\alpha}_A|^2+c_2|\alpha_{A_2}|^2+c_1\left(\overline{\tilde{\alpha}}_A\overline{\tilde{\alpha}}_A\tilde{\alpha}_{A^2}+c.c.\right)+(\overline{\tilde{\alpha}}_A+c.c.),
\label{eq:nondimham}
\end{equation}
with parameters 
\begin{eqnarray}  
t_0&=\frac{1}{E_A},\ \alpha_{0}=\frac{k_2}{E_A},\ c_1=\frac{k_1k_2}{E_A^2},\ c_2=\frac{E_{A_2}}{E_A}.
\end{eqnarray}  
Therefore, the dynamics of the system is completely determined by the choice of the two parameters $c_1,c_2\in\mathbb{R}$ and its initial conditions. Note that due to our choice of parameters $c_1=0$ implies $k_1=0$, i.e.\ no molecule formation, and arbitrary $k_2$. Moreover, the constraint stemming from the validity of the mean field approximation can now precisely be expressed as $c_1\ll 1$ which translates the constraint on the molecular reaction constant to $k_1\ll\frac{E_A^2}{k_2}$. 

What is the physical meaning of those two parameters? The ODE system \eref{eq:coupled} is equivalent to a pair of nonlinearly coupled harmonic oscillators. Whereas the oscillator describing the atoms has eigenfrequency one, $c_2$ determines the eigenfrequency of the molecular oscillator. Moreover, $c_1$ is the coupling strength between the two oscillators and is the only nonlinear term in the system. We therefore expect a regular behaviour for small values of $c_1$. Moreover, we see from \eref{eq:nondimham} that if $c_1\gg1$ the system is also integrable due to the conservation of the overall particle number $N_{tot}=|\tilde{\alpha}_A|^2+2|\tilde{\alpha}_{A_2}|^2$. In what follows, we investigate the different dynamical regimes determined by the choice of $c_1$ and fix $c_2$ to an experimentally realistic value.
\begin{figure}[htbp]
	\centering
		\includegraphics[width=0.8\textwidth]{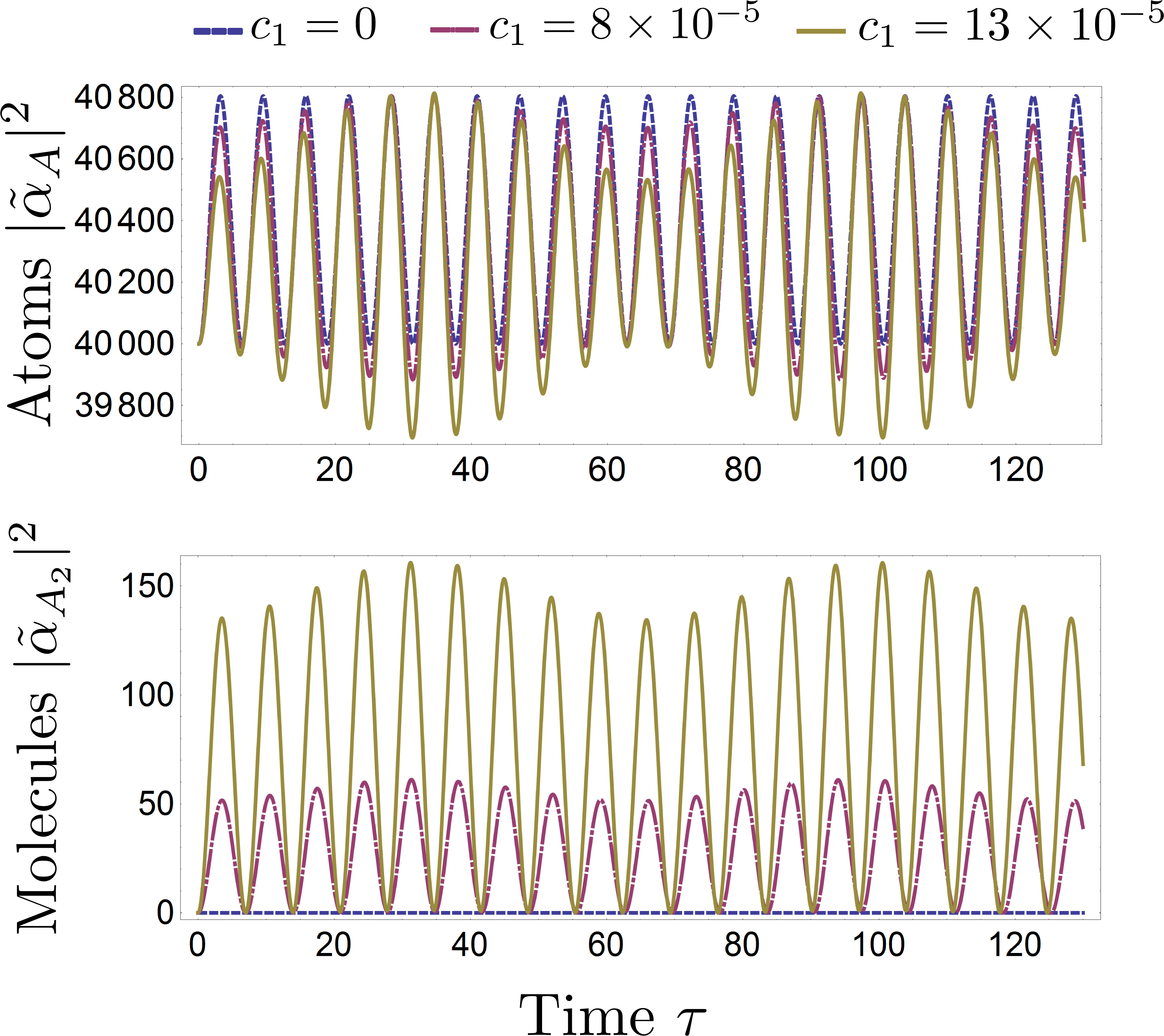}
	\caption{Time evolution of the (nondimensionalized) average number of atoms and molecules in mean field approximation. The relative energy of the ground states is $c_2=1.1$ and the initial number of atoms is $40000$. The oscillation of becomes more and more modulated with increasing molecule formation $c_1$. Perturbation theory determines an amplitude $A_{mod}\approx111$ atoms for $c_1=8\times10^{-5}$ and $A_{mod}\approx320$ atoms for $c_1=13\times10^{-5}$ (see \eref{eq:ammod}).}
	\label{fig:merged}
\end{figure}
Applying perturbation theory for anharmonic oscillators \cite{hinch1991perturbation} provides us analytic solutions of the occurring dynamical phenomena for sufficiently small coupling constants $c_1$: The trajectories depicted in figure \ref{fig:merged} show the expectation value of atoms $|\tilde{\alpha}_A|^2$ and molecules $|\tilde{\alpha}_{A_2}|^2$ for an initially depleted molecular mode and $40000$ atoms. In case of no coupling at all, i.e., $c_1=0$, the molecular mode remains completely depleted, whereas the atomic mode oscillates due to the coupling to the bath. However, as we increase $c_1$ the system progressively enters a \emph{modulational regime}, in which the molecular site regularly oscillates and the amplitude of the free oscillation on the atomic site is modulated. Let $A_0=\tilde{\alpha}_A(0)$ denote the square root of the initial nondimensionalized number of atoms, then we obtain the following analytical expressions for the amplitude $A_{mod}$ and frequency $\omega_{mod}$ of this modulation: 
\begin{eqnarray}\eqalign{
      A_{mod}&=\frac{4 (A_0+1) A_0^3 c_1 ^2}{(c_2 -2)^2}+\mathcal{O}\left(c_1^4\right), \\
 \omega_{mod}&=c_2-1+\mathcal{O}\left(A_0^2c_1^2\right).}
\label{eq:ammod}
\end{eqnarray}
This means increasing $c_1$ causes an quadratic increase of $A_{mod}$ whereas the frequency of modulation $\omega_{mod}$ remains approximately unchanged. The restriction for the parameters for perturbation theory to be valid are $c_1^2A_0^2\ll 1\ll A_0$ and $c_2\in\left(1,2\right)$.

\begin{figure}[htbp]%
\centering
\subfloat[][]{%
\label{fig:ex3-a}%
\includegraphics[width=0.25\textwidth]{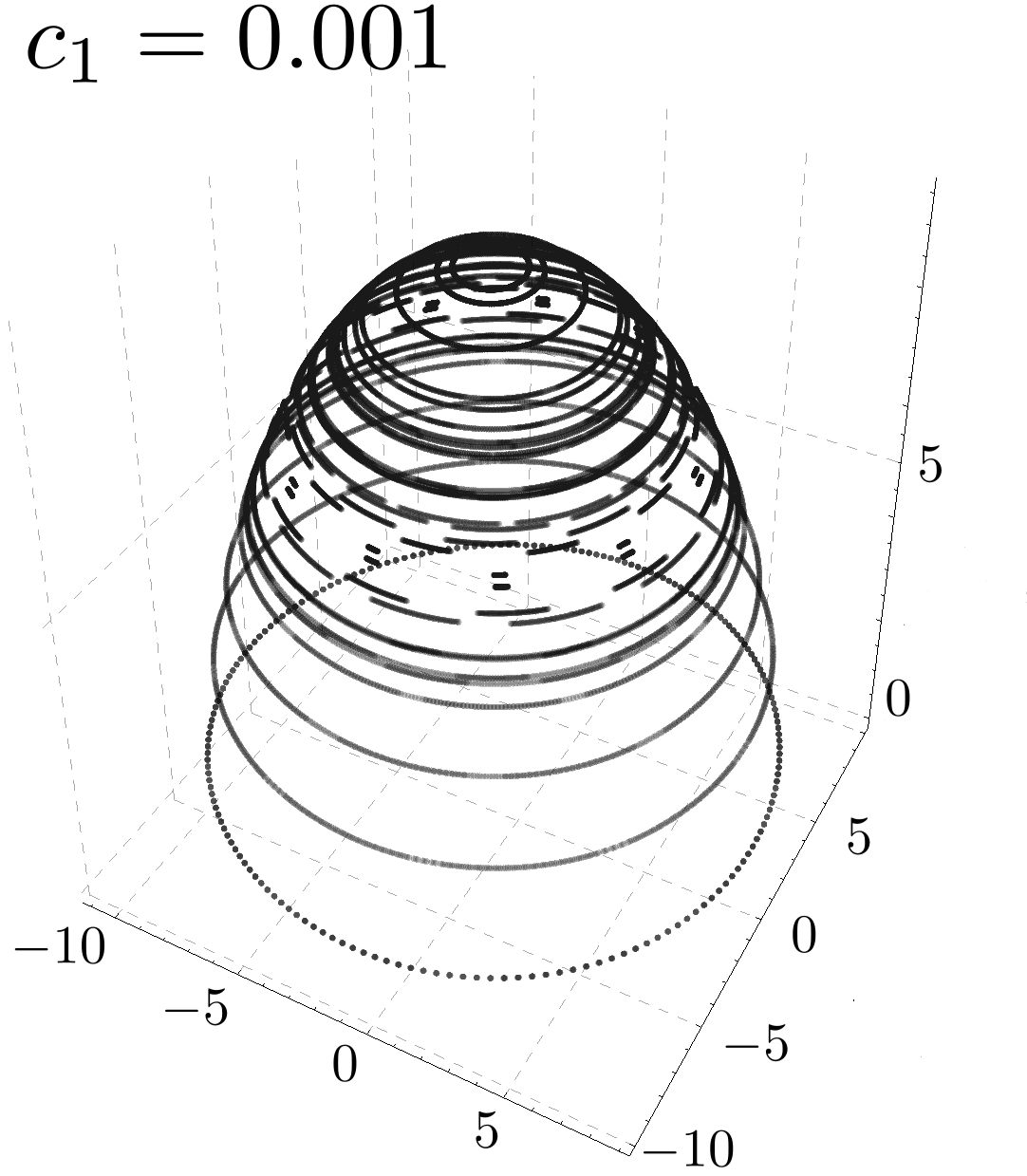}}%
\hspace{8pt}%
\subfloat[][]{%
\label{fig:ex3-b}%
\includegraphics[width=0.25\textwidth]{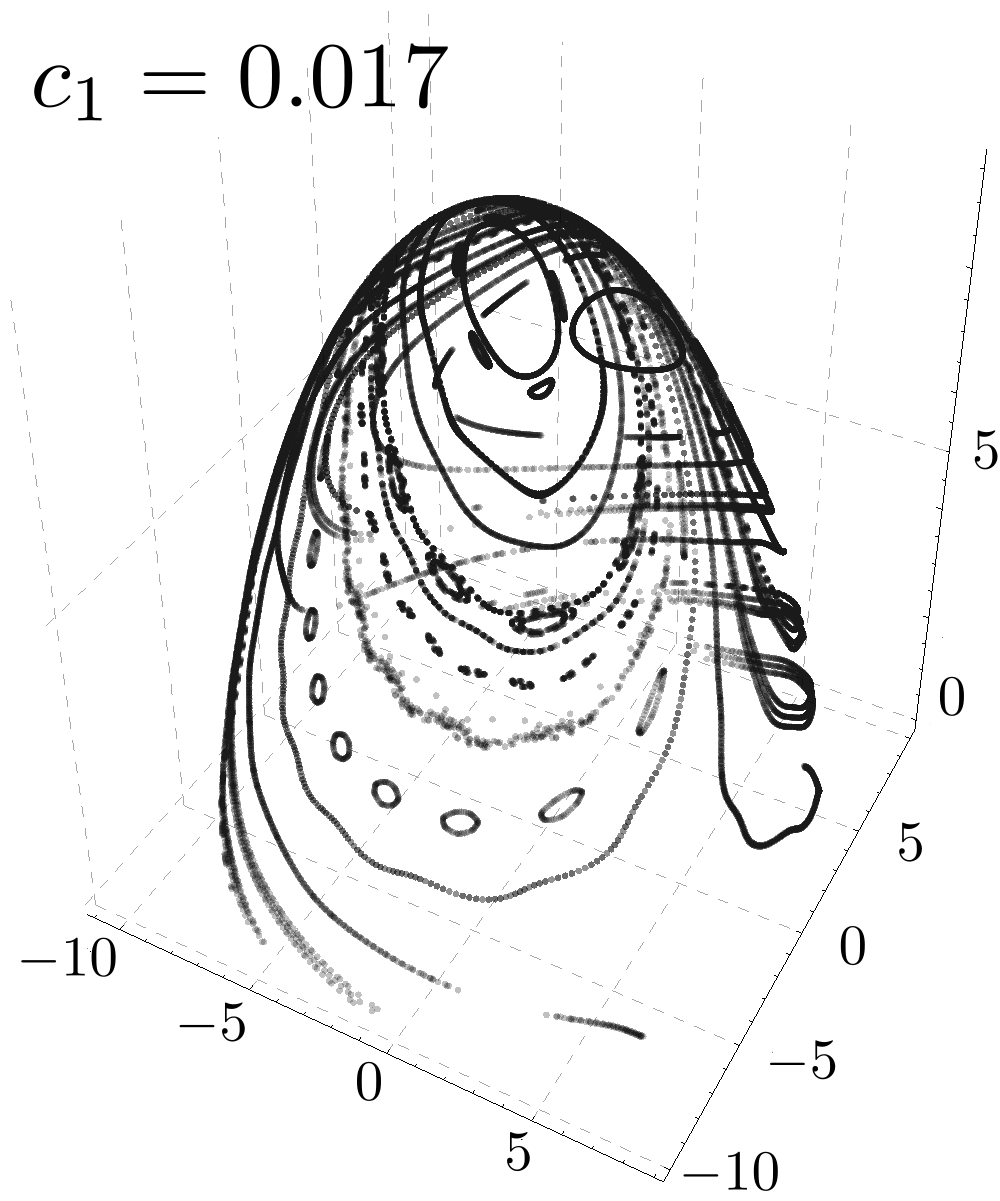}}\\

\subfloat[][]{%
\label{fig:ex3-c}%
\includegraphics[width=0.25\textwidth]{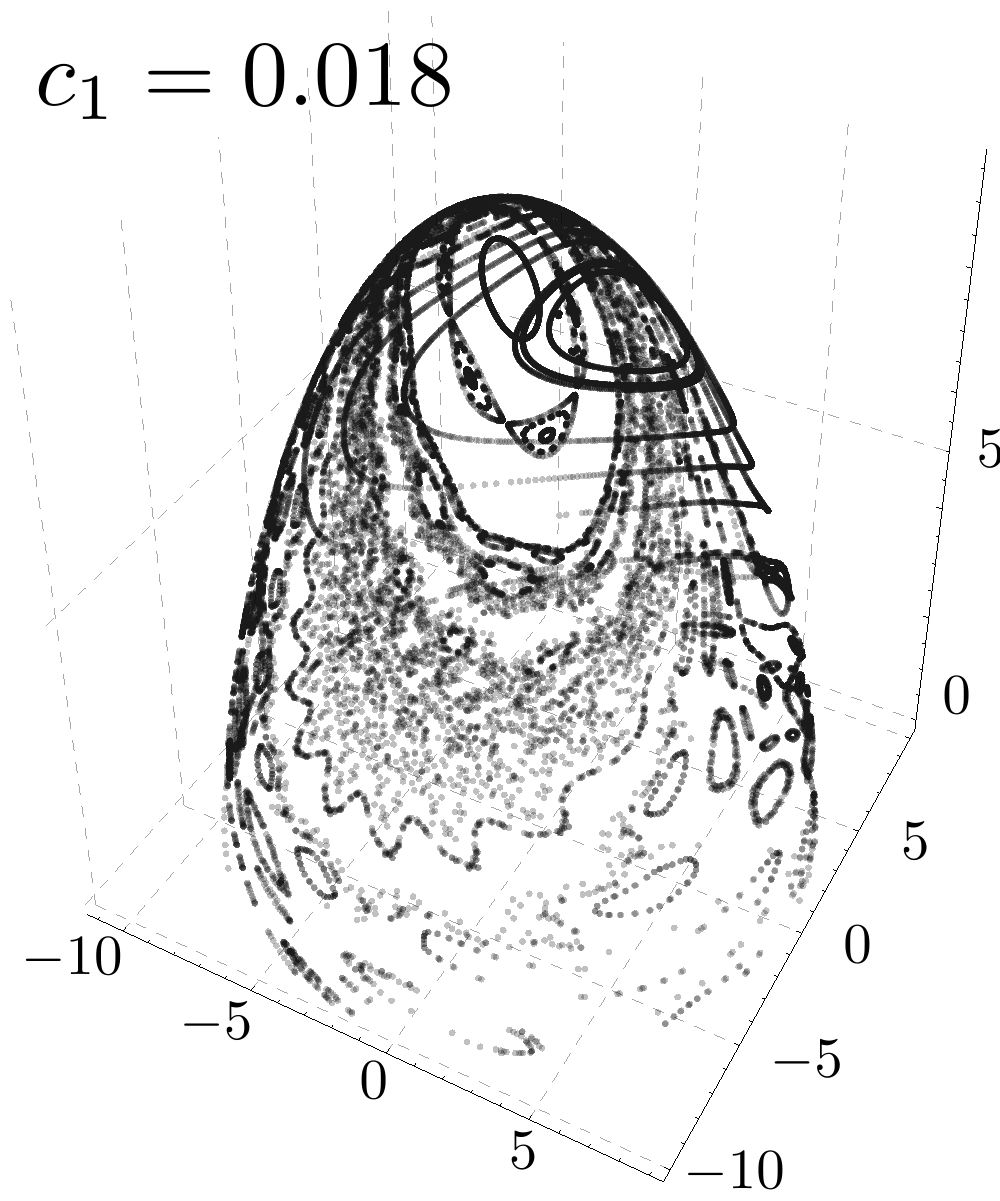}}%
\hspace{8pt}%
\subfloat[][]{%
\label{fig:ex3-d}%
\includegraphics[width=0.25\textwidth]{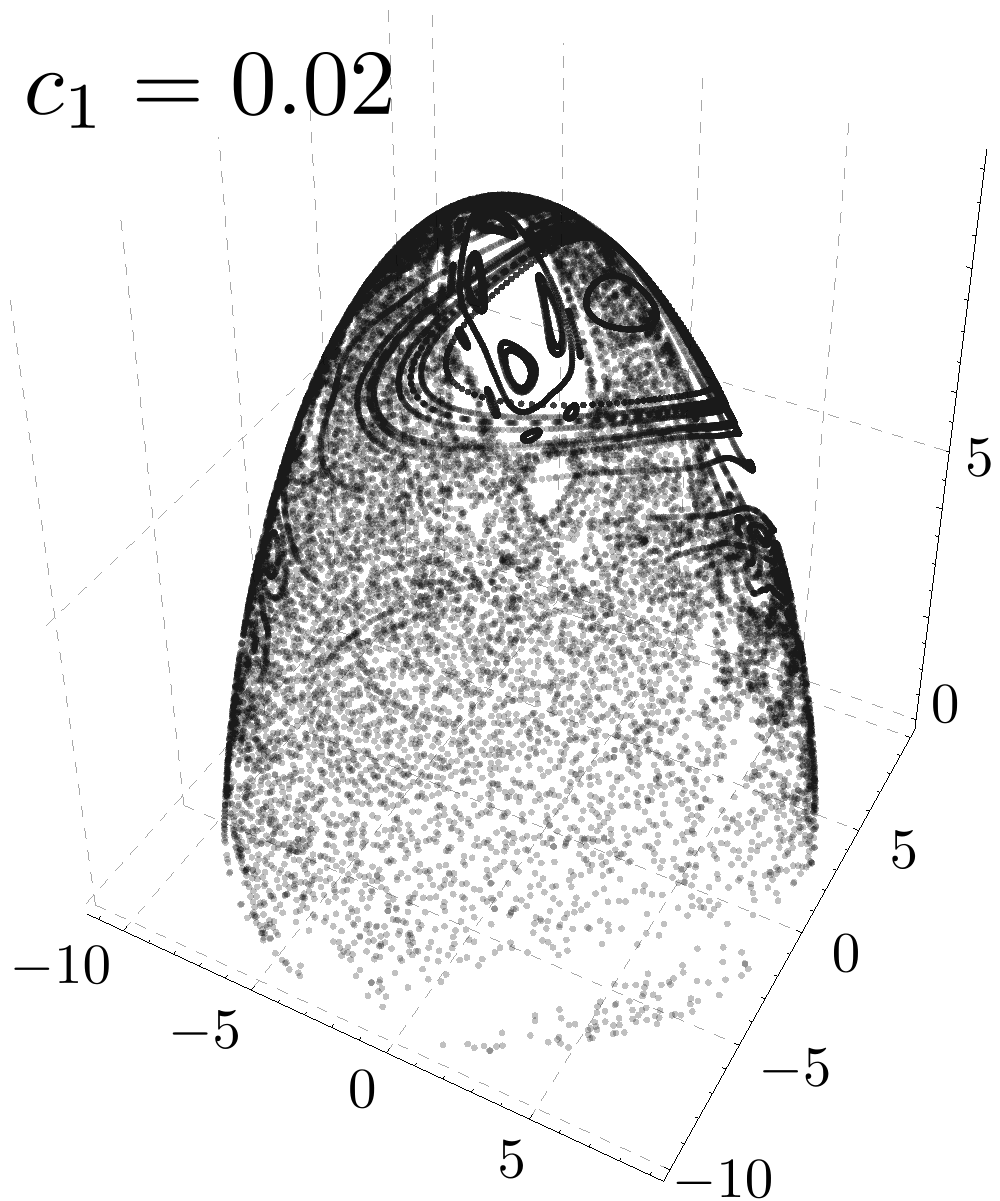}}\\

\subfloat[][]{%
\label{fig:ex3-e}%
\includegraphics[width=0.25\textwidth]{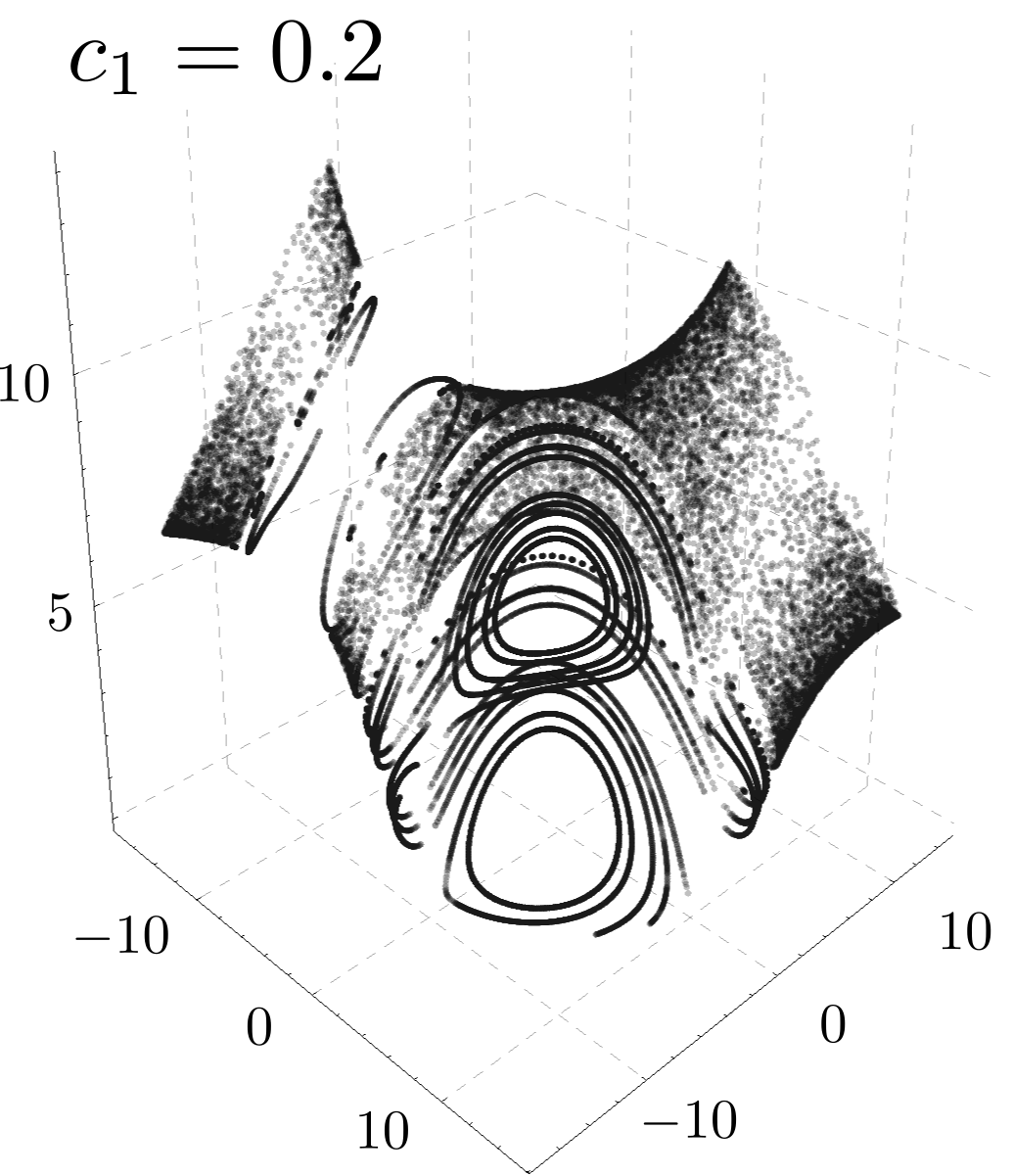}}%
\hspace{8pt}
\subfloat[][]{%
\label{fig:ex3-f}%
\includegraphics[width=0.25\textwidth]{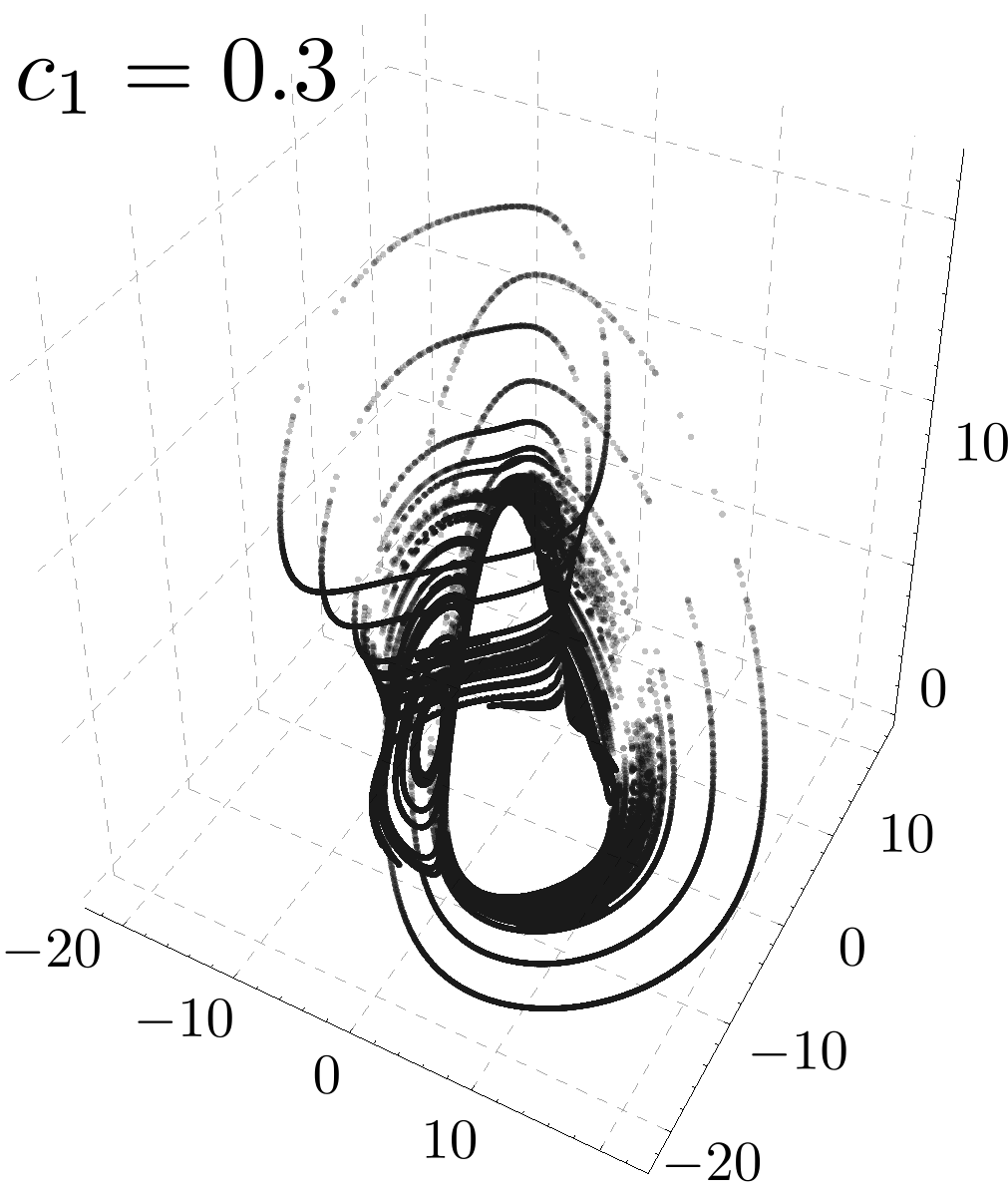}}%
\caption[Poincare Sections]{Poincar\'e sections with $X_A$ and $P_A$ on the (x,y)-axes. We set $c_2=1.1$ and the energy $E=100$. Note that in contrast to usual Poincar\'e sections in the literature, we plotted $P_{A_2}$ on the z-axis to get an better impression of the projected energy hypersurface. The sections are plotted for 25 long-time trajectories with arbitrary initial conditions: For a small perturbation pictured in \subref{fig:ex3-a} the system remains integrable. Increasing the perturbation in \subref{fig:ex3-b} leads to splitting of the first orbits into little islands according to the Poincar\'e-Birkhoff theorem \cite{poincare}. In \subref{fig:ex3-c} the irregular trajectories begin to spread out and densely fill out the energy hypersurface. In \subref{fig:ex3-d} most of the energy hypersurface is covered by irregular trajectories. Figure \subref{fig:ex3-e} shows a significant deformation of the energy hypersurface. In \subref{fig:ex3-f} all of the sampled trajectories are again on integrable curves.}%
\label{fig:poincare}%
\end{figure}

What happens to the system, if we increase $c_1$ beyond the regime of perturbation theory? We already mentioned that $c_1$ interpolates between integrable systems. But does the system remain integrable for all choices of $c_1$? A well-known tool to characterize irregular behaviour of a system is to consider its Poincar\'e sections \cite{ott2002chaos}. In our case, we choose the quadratures $X_i=\frac{1}{2}(\alpha_i+\overline{\alpha_i})$ and $P_i=\frac{1}{2i}(\alpha_i-\overline{\alpha_i})$ with $i\in\{A,A_2\}$ as dynamical variables and the surface $X_{A_2}=0$ as intersection surface. Illustrative examples of these sections for different $c_1$ are shown in figure \ref{fig:poincare}. We find that for a certain range of $c_1$ the system shows behaviour, which is typical for Hamiltonian chaos: As long as the system remains integrable, the Poincar\'e section consists of closed curves corresponding to sections of two-dimensional tori. However, increasing $c_1$ deforms and finally destroys some of the closed curves. Some of the sampled trajectories start to densely fill out parts of the energy hypersurface. We call this the \emph{chaotic regime} of the reaction. Finally, further increase of $c_1$ leads to deformation of the energy hypersurface and eventually restores the integrability of the system. 

\section{Summary and Outlook}
We have presented a proposal to systematically investigate the kinetics of ultracold chemical reactions. The dynamical analysis for the most elementary bosonic examples already implies that entanglement plays a major role in the formation of ultracold molecules. This leads to a relaxation of the atomic and molecular subsystems resulting in the stationarity of local observables similar to quenched systems. Moreover, considering two concurrent elementary reactions amounts to complex dynamical behaviour like Hamiltonian chaos in the mean field approximation. Extending our analysis to fermionic or mixed systems and validating the predicted phenomena against experimental data are the next natural steps to be taken.

\section*{Acknowledgments} Helpful discussions with J.\ Eisert, R.\ Werner, E.\ Schreiber, S.\ Schmidt, L.\ C.\ Venuti, B.\ Neukirchen, A.\ Werner and A.\ Milsted are most gratefully acknowledged. This work was supported by the ERC grants QFTCMPS, SIQS, and POLAR and by the cluster of excellence EXC 201 Quantum Engineering and Space-Time Research.
\section*{References}
\bibliography{coldchemicsbib}

\begin{thebibliography}{10}

\bibitem{wynar2000molecules}
R.~Wynar, R.~S. Freeland, D.~J. Han, C.~Ryu, and D.~J. Heinzen.
\newblock Molecules in a bose-einstein condensate.
\newblock {\em Science}, 287(5455):1016--1019, November 2000.

\bibitem{hutson2010ultracold}
J.~M. Hutson.
\newblock Ultracold chemistry.
\newblock {\em Science}, 327(5967):788--789, December 2010.

\bibitem{donley2002atom}
E.~A. Donley, N.~R. Claussen, S.~T. Thompson, and C.~E. Wieman.
\newblock Atom--molecule coherence in a bose--einstein condensate.
\newblock {\em Nature}, 417(6888):529--533, May 2002.

\bibitem{regal2003creation}
C.~A. Regal, C.~Ticknor, J.~L. Bohn, and D.~S. Jin.
\newblock Creation of ultracold molecules from a fermi gas of atoms.
\newblock {\em Nature}, 424(6944):47--50, July 2003.

\bibitem{ospelkaus2010quantum}
S.~Ospelkaus, K.-K. Ni, D.~Wang, M.~H.~G. de~Miranda, B.~Neyenhuis,
  G.~Quéméner, P.~S. Julienne, J.~L. Bohn, D.~S. Jin, and J.~Ye.
\newblock Quantum-state controlled chemical reactions of ultracold
  potassium-rubidium molecules.
\newblock {\em Science}, 327(5967):853--857, 2010.

\bibitem{heinzen2000superchemistry}
D.~J. Heinzen, R.~Wynar, P.~D. Drummond, and K.~V. Kheruntsyan.
\newblock Superchemistry: Dynamics of coupled atomic and molecular
  bose-einstein condensates.
\newblock {\em Phys. Rev. Lett.}, 84:5029--5033, May 2000.

\bibitem{goral2001multimode}
K.~G\'oral, M.~Gajda, and K.~Rza\ifmmode \mbox{\c{}}\else
  \c{}\fi{}\ifmmode~\dot{z}\else \.{z}\fi{}ewski.
\newblock Multimode dynamics of a coupled ultracold atomic-molecular system.
\newblock {\em Phys. Rev. Lett.}, 86:1397--1401, Feb 2001.

\bibitem{Santos2005classical}
G.~Santos, A.~Tonel, A.~Foerster, and J.~Links.
\newblock Classical and quantum dynamics of a model for atomic-molecular
  bose-einstein condensates.
\newblock {\em Phys. Rev. A}, 73:023609, Feb 2006.

\bibitem{carr2009cold}
L.~D. Carr, D.~{DeMille}, R.~V. Krems, and J.~Ye.
\newblock Cold and ultracold molecules: science, technology and applications.
\newblock {\em New Journal of Physics}, 11(5):055049, May 2009.

\bibitem{quemener2010strong}
G.~Qu{\'e}m{\'e}ner and J.~L. Bohn.
\newblock Strong dependence of ultracold chemical rates on electric dipole
  moments.
\newblock {\em Phys. Rev. A}, 81(2):022702, 2010.

\bibitem{idziaszek2010universal}
Z.~Idziaszek and P.~S. Julienne.
\newblock Universal rate constants for reactive collisions of ultracold
  molecules.
\newblock {\em Phys. Rev. Lett.}, 104(11):113202, March 2010.

\bibitem{connors1990chemical}
K.~A. Connors.
\newblock {\em Chemical Kinetics: The Study of Reaction Rates in Solution}.
\newblock John Wiley \& Sons, 1990.

\bibitem{belousov1959periodic}
B.~P. Belousov.
\newblock A periodic reaction and its mechanism.
\newblock {\em Ref. Radiats. Med.}, 1958:145, 1959.

\bibitem{zhabotinsky1964periodical}
A.~M. Zhabotinsky.
\newblock Periodical oxidation of malonic acid in solution (a study of the
  belousov reaction kinetics).
\newblock {\em Biofizika}, 9:306--311, 1964.

\bibitem{epstein1996nonlinear}
I.~R. Epstein and K.~Showalter.
\newblock Nonlinear chemical dynamics: Oscillations, patterns, and chaos.
\newblock {\em J. Phys. Chem.}, 100(31):13132--13147, January 1996.

\bibitem{baez2012course}
J.~C. Baez and B.~Fong.
\newblock Quantum techniques for studying equilibrium in reaction networks.
\newblock {\em arXiv:1305.4988}.

\bibitem{cramer2008exact}
M.~Cramer, C.~M. Dawson, J.~Eisert, and T.~J. Osborne.
\newblock Exact relaxation in a class of nonequilibrium quantum lattice
  systems.
\newblock {\em Phys. Rev. Lett.}, 100(3):030602, January 2008.

\bibitem{cramer2010quantum}
M.~Cramer and J.~Eisert.
\newblock A quantum central limit theorem for non-equilibrium systems: exact
  local relaxation of correlated states.
\newblock {\em New J. Phys.}, 12(5):055020, May 2010.

\bibitem{Mar-Sarao_08}
R.~Mar-Sarao and H.~Moya-Cessa.
\newblock Optical realization of a quantum beam splitter.
\newblock {\em Opt. Lett.}, 33(17):1966--1968, September 2008.

\bibitem{yurovsky2006formation}
V.~A. Yurovsky.
\newblock Formation of molecules and entangled atomic pairs from atomic {BEC}
  due to feshbach resonance.
\newblock {\em arXiv:cond-mat/0611054}.

\bibitem{walls1972nonlinear}
D.~F. Walls and C.~T. Tindle.
\newblock Nonlinear quantum effects in optics.
\newblock {\em J. Phys. A: Gen. Phys.}, 5(4):534, 1972.

\bibitem{hillery1990photon}
M.~Hillery.
\newblock Photon number divergence in the quantum theory of n-photon down
  conversion.
\newblock {\em Phys. Rev. A}, 42(1):498--502, July 1990.

\bibitem{javanainen1999coherent}
J.~Javanainen and M.~Mackie.
\newblock Coherent photoassociation of a bose-einstein condensate.
\newblock {\em Phys. Rev. A}, 59(5):R3186--R3189, May 1999.

\bibitem{javainen2}
M.~Ko\u{s}trun, M.~Mackie, R.~C\^ot\'e, and J.~Javanainen.
\newblock Theory of coherent photoassociation of a bose-einstein condensate.
\newblock {\em Phys. Rev. A}, 62:063616, Nov 2000.

\bibitem{santos_links}
G.~Santos, A.~Foerster, J.~Links, E.~Mattei, and S.~R. Dahmen.
\newblock Quantum phase transitions in an interacting atom-molecule boson
  model.
\newblock {\em Phys. Rev. A}, 81:063621, Jun 2010.

\bibitem{vardi2001quantum}
A.~Vardi, V.~A. Yurovsky, and J.~R. Anglin.
\newblock Quantum effects on the dynamics of a two-mode atom-molecule
  bose-einstein condensate.
\newblock {\em Phys. Rev. A}, 64(6):063611, 2001.

\bibitem{vardi2001bose}
A.~Vardi and J.~R. Anglin.
\newblock Bose-einstein condensates beyond mean field theory: Quantum
  backreaction as decoherence.
\newblock {\em Phys. Rev. Lett.}, 86:568--571, Jan 2001.

\bibitem{hines2003entanglement}
A.~P. Hines, Ross~H. {McKenzie}, and G.~J. Milburn.
\newblock Entanglement of two-mode bose-einstein condensates.
\newblock {\em Phys. Rev. A}, 67(1):013609, January 2003.

\bibitem{cramer2008exploring}
M.~Cramer, A.~Flesch, I.~P. {McCulloch}, U.~Schollwöck, and J.~Eisert.
\newblock Exploring local quantum many-body relaxation by atoms in optical
  superlattices.
\newblock {\em Phys. Rev. Lett.}, 101(6):063001, August 2008.

\bibitem{neumann1}
J.v. Neumann.
\newblock Beweis des ergodensatzes und des h-theorems in der neuen mechanik.
\newblock {\em Z. Phys.}, 57(1-2):30--70, 1929.

\bibitem{rigol2008thermalization}
M.~Rigol, V.~Dunjko, and M.~Olshanii.
\newblock Thermalization and its mechanism for generic isolated quantum
  systems.
\newblock {\em Nature}, 452(7189):854--858, 2008.

\bibitem{Srednicki}
M.~Srednicki.
\newblock The approach to thermal equilibrium in quantized chaotic systems.
\newblock {\em J. Phys. A}, 32(7):1163, 1999.

\bibitem{xie2009experimental}
F.~Xie, V.B. Sovkov, A.~M. Lyyra, D.~Li, S.~Ingram, J.~Bai, V.S. Ivanov,
  S.~Magnier, and L.~Li.
\newblock Experimental investigation of the $cs_2 \ a \ ^{3}\sigma_u^+$ triplet
  ground state: Multiparameter morse long range potential analysis and
  molecular constants.
\newblock {\em The Journal of chemical physics}, 130(5):051102, 2009.

\bibitem{krauss1990effective}
M.~Krauss and W.J. Stevens.
\newblock Effective core potentials and accurate energy curves for cs2 and
  other alkali diatomics.
\newblock {\em The Journal of Chemical Physics}, 93(6):4236--4242, 1990.

\bibitem{allouche2012transition}
A.-R. Allouche and M.~Aubert-Fr{\'e}con.
\newblock Transition dipole moments between the low-lying
  $\Omega_{g,u}^{(+/-)}$ states of the $Rb_2$ and $Cs_2$ molecules.
\newblock {\em The Journal of chemical physics}, 136(11):114302, 2012.

\bibitem{vexiau_optimal_2011}
R.~V\'{e}xiau, N.~Bouloufa, M.e Aymar, J.~G. Danzl, M.~J. Mark, H.-C. Naegerl,
  and O.~Dulieu.
\newblock Optimal trapping wavelengths of $Cs_2$ molecules in an optical
  lattice.
\newblock {\em arXiv:1102.1793}.

\bibitem{hinch1991perturbation}
E.~J. Hinch.
\newblock {\em Perturbation methods}.
\newblock Cambridge university press, 1991.

\bibitem{ott2002chaos}
E.~Ott.
\newblock {\em Chaos in dynamical systems}.
\newblock Cambridge university press, 2002.

\bibitem{poincare}
H.~Poincar\'e.
\newblock Sur un th\'eor\`eme de g\'eometrie.
\newblock {\em Rend. Circ. Mat. Palermo}, 33:375 – 407, 1912.

\end{thebibliography}
\bibliographystyle{unsrt}

\end{document}